\documentclass[aps,pre,10pt,twocolumn,superscriptaddress]{revtex4-2}
\usepackage{graphicx}
\usepackage{epstopdf}
\usepackage{xcolor}
\usepackage{amsmath}
\usepackage{amssymb}
\usepackage[utf8]{inputenc}
\usepackage{graphicx,float,hyperref}
\usepackage{natbib}
\usepackage{graphicx,latexsym} 

\begin{document}


\title{Optimization analysis of an  endoreversible quantum heat engine with efficient power function}

\author{Kirandeep Kaur}
\email{ kiran.nitj@gmail.com}
\affiliation{Department of Physical Sciences, Indian Institute of Science Education and Research Mohali, Sector 81, S.A.S. Nagar, Manauli PO 140306, Punjab, India}
\author{Anmol Jain}
\affiliation{  Department of IT,   
	Dr B R Ambedkar National Institute of Technology Jalandhar, Punjab-144027, India}
\author{Love Sahajbir Singh}
\affiliation{ Department of IT,   
	Dr B R Ambedkar National Institute of Technology Jalandhar, Punjab-144027, India}
\author{Rakesh Singla}
\affiliation{ Department of Physics,   
	Dr B R Ambedkar National Institute of Technology Jalandhar, Punjab-144027, India}
	\author{Shishram Rebari}
\email{rebaris@nitj.ac.in}
\affiliation{ Department of Physics,   
Dr B R Ambedkar National Institute of Technology Jalandhar, Punjab-144027, India}

%
%


\begin{abstract}
We study the optimal performance of an endoreversible quantum dot heat engine, in which the heat transfer between the system and baths is mediated by
 qubits, operating under the conditions of a trade-off objective function known as maximum efficient power function defined by the product of power and efficiency of the engine.  First, we numerically study the optimization of the efficient power function for the engine under consideration. Then, we obtain some analytic results by applying high-temperature limit and compare the performance of the engine at maximum efficient power function to the engine operating in the   maximum power regime. We find that the engine operating at maximum efficient power function  produces at least 88.89\% of the maximum power output while at the same time reduces the power loss due to entropy production  by considerable amount. We conclude by studying the stochastic simulations of the efficiency of the engine in maximum power and maximum efficient power regime. We find that the engine operating at maximum power is subjected to less power fluctuations  as compared to the on one operating at maximum efficient power function.
\end{abstract}


\maketitle 

\section{Introduction}

Back in 1824, Sadi Carnot proposed an ideal thermodynamic cycle, known as Carnot cycle, working between two reservoirs at different temperatures $T_c$ and $T_h$ ($T_c<T_h$) to convert  heat into work. The efficiency of Carnot cycle, $\eta_C=1-T_c/T_h$, is a universal result.  However, due to its reversible nature, it takes infinite time to complete one Carnot cycle, thereby producing vanishing power output per cycle. 
However, realistic engines should work under irreversible conditions by taking into account the finite-time constraints and produce finite-power output \cite{CA1975,Andresen2011,Berry1984}.  In search for optimizing the performance of realistic heat engines, the field of finite-time thermodynamics was developed \cite{Salamon2001,Berry1984,Andresen2011,devosbook,Berrybook}.  Curzon-Albohm (CA) were the pioneers of the finite-time thermodynamic and they introduced a simple Carnot-like heat engine model, known as endoreversible model \cite{CA1975,Rubin}, in which irreversible heat transfer between the working medium and the reservoirs is assumed to obey Newton's law of conduction \cite{CA1975}. The efficiency of endoreversible at maximum power is given by
\begin{equation}
\eta_{\rm CA} = 1- \sqrt{1-\eta_C}=  \frac{\eta_C}{2}+\frac{\eta_C^2}{8}+\frac{\eta_C^3}{16} + ....... 
\end{equation}
As in the case of the Carnot efficiency, the  CA efficiency  depends on the ratio of reservoir temperatures only. The endoreversible model has paramount importance in the field of finite-time thermodynamics as many different models of heat engines, working under the conditions of maximum power \cite{Tu2008,Esposito2010,VarinderJohal2018,Schmiedl2008,Lutz2012,Geva1992,Kosloff1984}, share the universality of efficiency up to the second order term in $\eta_C$ \cite{Esposito2009,Tu2008}  with CA efficiency.

Over the past few decades, the technological advanced made it possible to manipulate quantum systems operating at nanoscale \cite{Uzdin2019,Josefsson:2018,Peterson2019PRL,VV2022,Scarani2019,Rossnagel2016}.  Since the size and quantum effects play an important role in thermodynamic properties of the system, the extension of  the theory of classical thermodynamics to the quantum
systems is called for \cite{Sai2016,Sourav2021,Asli2020,AlickiKosloff,DeffnerBook,Mahler}. Quantum heat engines, being the technological devices of practical importance, provide us with ideal platform to study the relation between classical and quantum thermodynamics \cite{LevyKosloff,Benenti2017,MyersReview}.  

Further, the rapid development in the field of quantum technologies has bring up the question of resource consumption in the thermodynamic landscape \cite{Alexia2022}. The heat engines working at maximum power regime are also known to waste a large amount of fuel due to large amount of entropy production, which pollutes the environment in turn \cite{Chen2001,devosbook}.  Thus taking into account the ecological and economical concerns, one should operate heat engines in such a regime which  establishes a compromise between the efficiency and power of the heat engine  \cite{Chen2001,devosbook,ABrown1991,VJ2019,VJ2017,Kiran2021,VS2022,Hernandez2001,Arias2009,VJ2020}. Efficient power function is such an alternative trade-off objective function which is defined by the product of efficiency and power of the heat engine \cite{Stucki}, thus taking caring of the compromise between power and efficiency. It was introduced by Stucki to study the biochemical energy conversion process \cite{Stucki}. Over the past three decades, it has been successfully used in  studying the energy conversion process in classical \cite{Yilmaz,YanChen1996,VJ2018,Me5,Zhang2017,Chimal,Sanchez2016,Kumari2020} as well as quantum thermal devices \cite{MePRR,Deffner2020}.   
 
In this work, we study the optimal performance of a quantum endoreversible model \cite{Du2020,Fernandez2022}, in which heat transfer between the working fluid and the reservoirs is mediated by qubits, working under the conditions of maximum efficient power function. Unlike the classical endoreversible heat engine, the laws of heat transfer no longer obey the Newton's law of conduction and filtration of the heat current through the quantum qubits introduces the quantum effects in the performance of the quantum version of endoreversible  engine. 

The paper is organized as follows. In Sec. I we introduce the model of quantum endroreversible heat engine. In Sec III we numerically optimize  the performance of the engine and present our result. In Sec IV, we restrict ourselves to the high-temperature limit and  compare the performance of the heat engine at maximum efficient power to the engine  at maximum power.  Sec. V is devoted to study of the stochastic fluctuations in the power output of the the engine. We conclude in Sec. VI.
 
			\section{ Model and theory}	
		\begin{figure}
			\begin{center}
				\includegraphics[width=8.7cm]{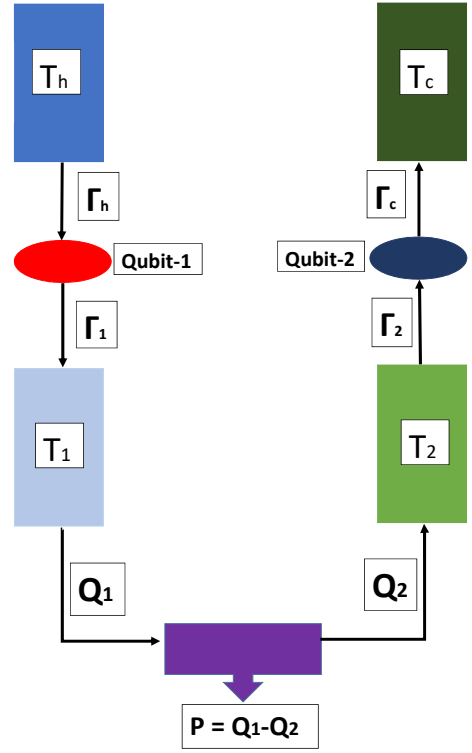}
			\end{center}
			\caption{The schematic diagram of quantum dot heat engine with irreversible heat transfer.  In this model we have two qubits (1 and 2) which interact with the two  bath. The location of qubit 1 is in between the hot bath ($T_{h}$) and working medium. The temperature of qubit 1 is denoted  by  $T_{1}$. In the similar way qubit 2 is located in between the cold bath ($T_{c}$) and working medium. The temperature of qubit 2 is denoted  by  $T_{2}$. The coupling strength of qubit 1  with hot bath and working medium is,  $\Gamma_{h}$  and $\Gamma_{1} $ respectivly. In the similar way the coupling strength of qubit 2  with cold  bath and working medium is    $\Gamma_{c}$  and $\Gamma_{2} $ respectivly. $E_{1} $ and $E_{2} $ represents the energy level value of qubit 1 and qubit 2 respectively. }
		\end{figure}
The schematics of the  quantum version of the  endoreversible engine is shown in Fig. 1. During the hot isothermal branch , the qubits 1, with  energy gap $E_1$,  induces the irreversible heat flux $Q_1$  from  the thermal bath at temperature $T_h$ to the working fluid at temperature $T_1$. Similarly, during the cold isothermal branch,  heat flux $Q_2$ from the working fluid at temperature $T_2$ to the cold reservoir at temperature $T_c$ is mediated via  qubit 2 having energy gap $E_2$. The model in Fig. 1 will operate as an heat engine provided the following condition is satisfied: $T_h \geq T_1\geq T_2\geq T_c$.
 
The hamiltonian of the   k ($k=1, \,2$) qubit  is given as:
		\begin{equation}
			\hat{H}_{k}= \frac{E_{k}}{2} \hat{\sigma}_{k}^{z},
		\end{equation}
		and 
		 
	 	where $ \sigma_{k}^{z}$ is third component of spin-$\frac{1}{2}$ Pauli matrix. The dynamics of the qubit 1 (2) is governed by the following master equation		
		\begin{equation}
			\dot{\rho}_{1 (2)} = -i [\hat{H}_{1 (2)},\rho_{1 (2)}]+D_{h (c)}[\rho_{1 (2)}]+D_{1 (2)}[\rho_{1 (2)}],
		\end{equation}
		where the $D_{h (c)}[\rho_{1 (2)}]$ and $D_{1 (2)}[\rho_{1 (2)}]$ are the Lindblad superoperators associated with  baths of temperature of $T_{h (c)}$ and $T_{1 (2)}$,  respectively. The form  of these Lindblad dissipators for the qubit 1 is given by the following equation:
		\begin{eqnarray}
			D_{i}[\rho_{1}] &=& \Gamma_i  (n_{i}+1)\left[\sigma_{1}^{-}\rho_{1}\sigma_{1}^{+}-\frac{1}{2}\Big\{\sigma_{1}^{+}\sigma_{1}^{-},\rho_{1}\Big\}\right]\nonumber
			\\
			&&+\,  \Gamma_{i} n_{i}\left[\sigma_{1}^{+}\rho_{1}\sigma_{1}^{-} -\frac{1}{2} \Big\{\sigma_{1}^{-} \sigma_{1}^{+},\rho_{1}\Big\}\right],
		\end{eqnarray}
		where $i=1, h$ denote baths of temperature $T_{1}$ and $T_{h}$, respectively. $\Gamma_{i} $ denotes the dissipation rate associated with bath   i.  %
		 Similarly, for the qubit 2, we have
		\begin{eqnarray}
			D_{j}[\rho_{2}] &=& \Gamma_j (n_{j}+1)\left[\sigma_{2}^{-}\rho_{2}\sigma_{2}^{+}-\frac{1}{2}\Big\{\sigma_{2}^{+}\sigma_{2}^{-},\rho_{2}\Big\}\right]\nonumber
			\\
			&&+\,  \Gamma_{j} n_{j}\left[\sigma_{2}^{+}\rho_{2}\sigma_{2}^{-} -\frac{1}{2} \Big\{\sigma_{2}^{-} \sigma_{2}^{+},\rho_{2}\Big\}\right],
		\end{eqnarray}
		where $j=2,\,c$ for baths of temperature $T_{2}$ and $T_{c}$ respectively. $\sigma_{k}^{+} = |1><0|$ and 	$\sigma_{k}^{-}= |0><1| $ are raising and lowering operators for the qubit k ($k=1,\,2$), respectively. $n_{i (j)} = 1/ (e^{E_1 (2)/k_B T_i (j)})$ is the number of photons in the reservoir i (j) with energy gap $E_1 (2)$, where $i=1, h$ and $j=2, c$. From now on, we put $k_{B}=1$.

In the steady state, the populations of qubit 1 can be obtained by setting $\dot{\rho}_1=0$ in the left hand side of Eq. (3). Thus, we have		 
		\begin{equation}
			\rho_{1}^{s} =\frac{1}{2}(1+a_{1}^{z}\sigma_{1}^{z}),
		\end{equation}
		where $a_{1}^{z} = -(\Gamma_{h}+\Gamma_{1})/(\Gamma_{1} (2 n_1+1)+\Gamma_h (2 n_h + 1))$.
 In steady state, the heat flux $Q_{h} =Tr({H_{1}D_{h}[\rho_{1}^{s}]})$  flowing  out of the heat bath at $T_{h}$  is equal to the heat flux $Q_{1} =-Tr({H_{1}D_{1}[\rho_{1}^{s}]})$ entering bath 1 at temperature $T_1$. The final expression for the heat flux $Q_1$ absorbed by the engine is given by
		\begin{equation}
			Q_{1}= \gamma_{1}E_{1}(n_{h}-n_{1}),  \label{q1}
		\end{equation}
where $\gamma_1=(\Gamma_{h}+ \Gamma_{1})/(\Gamma_{1} (2 n_1+1)+\Gamma_h (2 n_h + 1))$.
	Similarly, the expression for the heat flux $Q_2$ rejected to cold reservoir at temperature $T_c$	is given by		
		\begin{equation}
			Q_{2}= \gamma_{2}E_{2}(n_{2}-n_{c}), \label{q2}
		\end{equation}
		 where $\gamma_2=(\Gamma_{c} \Gamma_{2})/(\Gamma_{2} (2 n_2+1)+\Gamma_c (2 n_c + 1))$.

		By using Eqs. (\ref{q1})	and . (\ref{q2}), the we can obtain  expressions for the power output and efficiency of the engine:	
\begin{eqnarray}
P &=& Q_1 - Q_2, \nonumber 
\\
&=& \gamma_1 E_1 \Big[ \frac{1}{e^{E_1/T_h}-1} -  \frac{1}{e^{E_1/T_1}-1}\Big]  \nonumber
\\
&& -  \gamma_2 E_2 \Big[ \frac{1}{e^{E_2/T_2}-1} -  \frac{1}{e^{E_2/T_c}-1}\Big], \label{power}
\end{eqnarray}
and
\begin{equation}
\eta = \frac{P}{Q_1} = 1 - \frac{\gamma_2 E_2 (n_1-n_c)}{\gamma_1 E_1 (n_h-n_1)}. \label{efficiency}
\end{equation}		
Further, for an endoreversible heat engine, the entropy productions of the working medium in two isothermal processes satisfy the following relation
	\begin{equation}
			\frac{Q_{1}}{T_{1}}=\frac{Q_{2}}{T_{2}},   \label{endocond}
		\end{equation}  
and 	the efficiency of the engine depends on internal temperatures of the working fluid during the isotherms:
			\begin{equation}
			\eta = 1- \frac{T_{2}}{T_{1}}. \label{efficiency2}       
		\end{equation}

B using Eqs.  (\ref{efficiency}),  (\ref{endocond}) and  (\ref{efficiency2}),  we get the relations between the  temperatures  $T_{1}$ and $T_{2}$ and the efficiency ($\eta$) as follows (see Ref. \cite{} for details)
		\begin{equation}
			T_{1}= \frac{E_{1}}{\ln\Big[\frac{1}{n_{h}-\frac{Q_{2}}{\gamma_{1}E_{1}(1-\eta)}}+1\Big]}, \quad
			T_{2}=  \frac{E_{2}}{\ln\Big[\frac{1}{n_{c}+\frac{Q_{1}(1-\eta)}{\gamma_{2}E_{2}}}+1\Big]}.
		\end{equation}
		
\section{optimization of  efficient power function }	
The optimization of power output of the quantum endoreversible engine has already been studied in Ref. \cite{Du2020}. Here, we would like  to study the optimal performance of endoreversible quantum heat engine operating at maximum efficient power function, which is more suitable objective function to optimize we if want to operate our engine in such a regime which pay equal attention to both efficiency and power of the engine. The expression for the efficient power function  ($P_\eta=\eta P$),  simply given by the product of power and efficiency of the engine, can be obtained by combining Eqs. (\ref{power}), 
(\ref{efficiency}), (\ref{endocond}) and (\ref{efficiency2}), and is given by
\begin{equation}
			P_{\eta} = \frac{\gamma_{2} E_{2} \eta^{2}}{1-\eta}  
	 \left\{         
			 \frac{1}
			{
			      \left[
			                  \frac{1}{n_h-P/(\gamma_1 E_1\eta)}
			      \right]^{     E_2/E_1(1-\eta)            }    -1 
			 }     -n_c
	 \right\}.  \label{EP}
		\end{equation}
Now, we will   optimize the efficient power function given in Eq. (\ref{EP}) by using numerical techniques.
\begin{figure}
			\begin{center}
				\includegraphics[width=8.6cm]{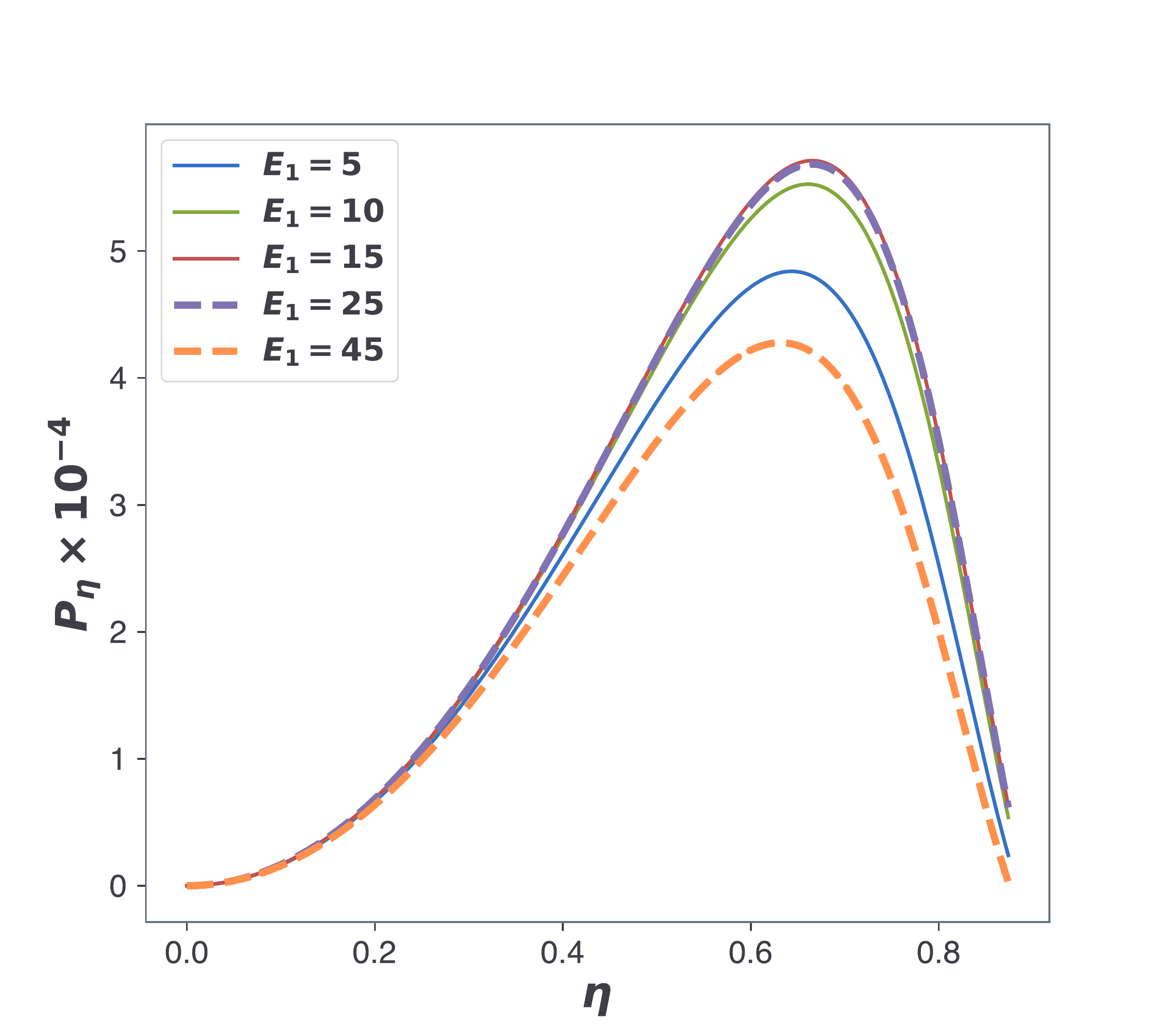}
			\end{center}
			\caption{Plot of the efficient power function (Eq. (\ref{EP}))  as a function of efficiency ($\eta$) of the engine for different set of values of   $E_{1}=5,\,10,\,15,\,25, \,\text{and} \,45$. The other parameters are fixed at constant values. $E_2=6$, $\Gamma_{h}=\Gamma_{1}=0.01$, $\Gamma_{c}=\Gamma_{2}=0.001$, 	 $T_{h}=10$, $T_{c}=1$. } 
		\end{figure}
	\begin{figure}
		\begin{center}
			\includegraphics[width=8.6cm]{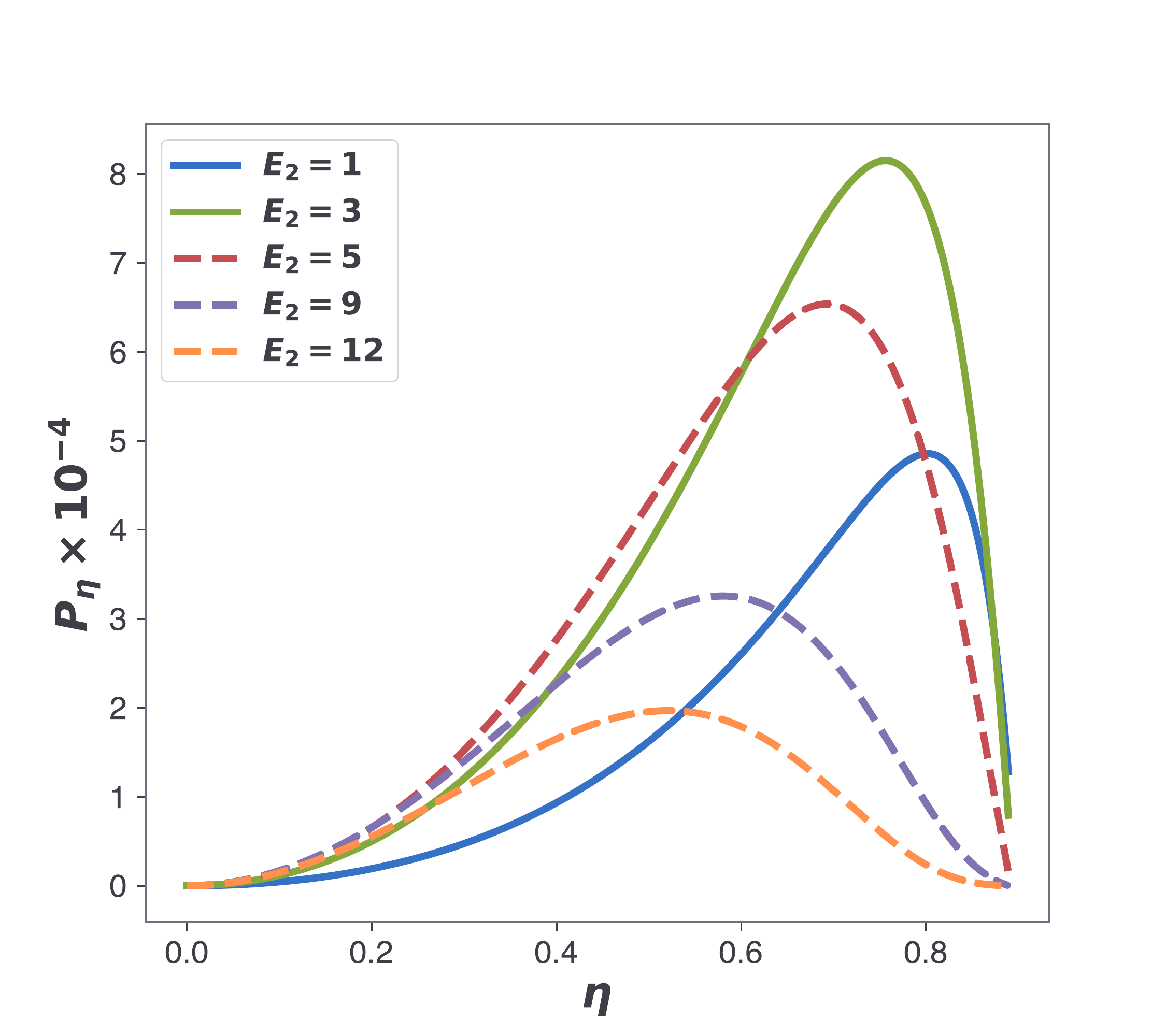}
		\end{center}
		\caption{Plot of the efficient power function (Eq. (\ref{EP}))  as a function of efficiency ($\eta$) of the engine for different set of values of   $E_{2}=1,\,3,\,5,\,9, \,\text{and} \,12$. The other parameters are fixed at constant values. $E_2=6$, $\Gamma_{h}=\Gamma_{1}=0.01$, $\Gamma_{c}=\Gamma_{2}=0.001$, 	 $T_{h}=10$, $T_{c}=1$. } 
	\end{figure}
Figs. 2 and 3 show the efficient power function varying with efficiency $\eta$ for different sets of values of  $E_{1}$ and  $E_{2}$, respectively. The trends in both figures are same. It is observed that if we increase the value of  $E_{1}$ and  $E_{2}$  beyond a certain value, then the efficient power function decreases very fast. This can be explained by looking into the expressions for average number of  photons with energy $E_1$ and $E_2$ in the reservoirs, which decreases with increasing energy gap $E_{1 (2)}$. Then, it can be seen from Eqs. (7) and (8) that power output of the engine must be very small for very large $E_1$ or $E_2$. On the other extreme, Eq. (\ref{power}) indicates that power output will be vanishingly small for very small values of $E_1$ and $E_2$. Thus, in order to optimize the performance of the engine, we have to tune  the energy gaps ($E_1$ or $E_2$) somewhere in between  the lower and higher values.  Thus, we can claim that performance of the endoreversible quantum engine is affected by the properties of the qubits, which controls the irreversible heat transfer between the reservoirs and the working fluid. By adjusting the energy gaps of the qubits, we can regulate power and efficiency of the engine. 
  
 In order to compare the performance of the engine operating at maximum efficient power function  to the engine at maximum power, we plot Fig. 4. It is clear from Fig. 4 that the engine at maximum efficient power yields larger efficiency that the engine operating at maximum power. This is due to the fact that optimization of efficient power function, by definition, takes care of the trade-off between efficiency and power of the engine, hence yielding more efficiency but slightly smaller power output than the engine operating at maximum power. The numerical values of maximum power and maximum efficient power function are given by $ P^{max}=8.76  \times  10^{-4} $ and $ P_{\eta}^{max}=5.53  \times  10^{-4}$, respectively. Corresponding  efficiencies at maximum power and efficient power functions are given by $\eta^{P}$= $0.593 $ and  $\eta^{P_{\eta}}$=$0.661$, respectively.		\begin{figure}
			\begin{center}
				\includegraphics[width=8.6cm]{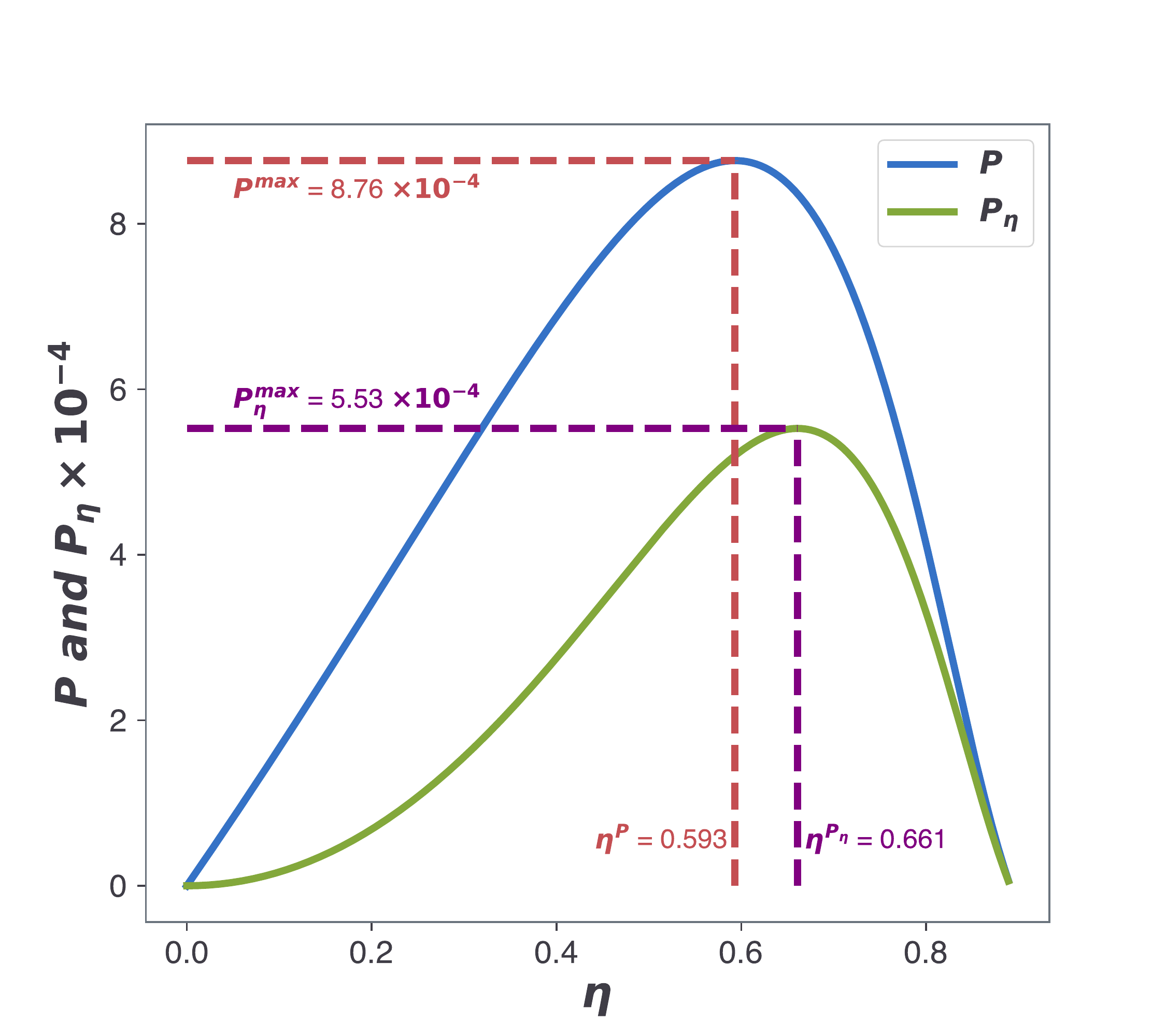}
			\end{center}
			\caption{Figure represents the comparison in between power curve and efficient power optimization varying with the efficiency $\eta$.  Both the curve has plotted for $E_{1}=10$ and  $E_{2}=6 $, rest of the parameters are, $\Gamma_{h}=\Gamma_{1}=0.01$, $\Gamma_{c}=\Gamma_{2}=0.001$, $T_{h}=10$, $T_{c}=1$. } 
		\end{figure}

		\section{ High Temperature Regime }
	Up to now, we have investigated the optimal performance of the engine by numerical techniques only. In order to obtain some analytic results, we 	will work in the  high-temperature limit (classical limit). In order to compare the performance of  quantum devices with the classical ones, is common to practice to work in the high-temperature limit \cite{Geva1994,VarinderTUR,Geva1996,Gevakosloff1992,VJ2019,VJ2020,Lutz2012,VOzgur2020,LinChen2003}.  In this case  temperature  of system is large as  compare to energies $E_{1} $ and $ E_{2}$, i.e. $\beta_{l}E_{i}\ll 1$ or $E_{i} \ll T_{l}$, 	$l = h,1,c,2$ and $i=1,2$. Similar to our previous consideration about the coupling between the thermal baths, we take $\Gamma_{1}=\Gamma_{h}$ and 	$\Gamma_{2}=\Gamma_{c}$. Applying these constraints, we can obtain the following expressions for  $Q_{1} $and $Q_{2}$
			\begin{equation}
			Q_{1} = \frac{\Gamma_{1}}{2(T_{1}+T_{h})}E_{1}(T_{h}-T_{1}) \label{auxheat1}
		\end{equation}
		\begin{equation}
			Q_{2} = \frac{\Gamma_{2}}{2(T_{2}+T_{c} )}E_{2}(T_{2}-T_{c}) \label{auxheat2}
		\end{equation}
Further, when $	T_{h}-T_{c} \ll T_{c} $, $T_{h}+T_{1}= 2T_{h}$ and $T_{c}+T_{2}= 2T_{c}$, Eqs. (\ref{auxheat1}) and  (\ref{auxheat2}) take the following forms
\begin{eqnarray}
Q_{1} &=& k_{1}(T_{h}-T_{1} ),  \label{auxheat11}
\\
Q_{2} &=& k_{2} (T_{2}-T_{c}),  \label{auxheat22}
\end{eqnarray}
where $k_{1}= \Gamma_{1}E_{1}/4 T_{h}$ and $k_{2} = \Gamma_{2} E_{2}/4T_{c}$. Both parameter $ r_{1}$ and $r_{2}$ directly depends on coupling parameteras, reservoirs temperatures and the energy gaps of the qubits. Thus in the high-temperature limit, the irreversible heat transfer between the reservoirs and the working fluid is governed by Newton's heat transfer law, just as in the case classical endoreversible heat engine.  

Using Eqs. (\ref{endocond}) and (\ref{efficiency2}) and eliminating $T_1$ and $T_2$ in Eqs.  (\ref{auxheat11}) and  (\ref{auxheat22}), we arrive at the following expressions for the power output, $P=\eta Q_h$, of the engine 
		\begin{equation}
			P^{\rm HT} = \frac{k_{1} k_{2}}{k_{1}+k_{2}} \eta \left(T_{h}-\frac{T_{c}}{1-\eta} \right). \label{auxxpower}
		\end{equation}
The optimization of Eq. (\ref{auxxpower}) with respect to $\eta$ yields the famous Curzon-Ahlborn efficiency:
\begin{equation}
\eta_{\rm CA} = 1-\sqrt{\frac{T_c}{T_h}} = 1-\sqrt{1-\eta_C} = \frac{\eta_C}{2} + \frac{\eta_C^2}{8} +   \frac{\eta_C^3}{16} + ..., \label{seriesCA}
\end{equation}
where we have used the relation $\tau=T_c/T_h=1-\eta_C$. Thus, in the high-temperature limit, our quantum endoreversible heat engine operates with the same efficiency at which classical endoreversible heat engine operates. This is expected result as high-temperature limit is considered to be classical limit and in this limit, many models of quantum heat engines and refrigerators operate at CA efficiency \cite{Geva1994,Geva1996,Gevakosloff1992,VJ2019,VJ2020,Lutz2012,LinChen2003}. 
	
Here, our motivation is to obtain some analytical result in order to compare the performance of the heat engine operating in the maximum power regime to one operating at maximum efficient power function. The expression for the efficient power function can be simply obtained by multiplying Eq. (\ref{auxxpower}) with efficiency $\eta$ of the engine,
\begin{equation}
P^{\rm HT}_\eta =\eta  P^{\rm HT} = \frac{k_{1} k_{2}}{k_{1}+k_{2}} \eta^2 \left(T_{h}-\frac{T_{c}}{1-\eta} \right). \label{auxxEP}
\end{equation}
Optimization of Eq. (\ref{auxxEP}) with respect to $\eta$ yields the following expression for the efficiency at maximum efficient power function		
\begin{eqnarray}
\eta^{P_\eta} &=& 1 - \frac{1}{4}(1-\eta_C)\left( 1+ \sqrt{1+\frac{8}{1-\eta_C}} \right)  \nonumber
\\
& = & \frac{2\eta_C}{3} + \frac{2\eta_C^2}{27} + \frac{10\eta_C^3}{243} +..... , \label{seriesYC}
\end{eqnarray}
which is exactly same as the efficiency at maximum efficient power of the classical endoreversible \cite{YanChen1996} and low-dissipation \cite{VJ2018} models of heat engine. By comparing Eqs. (\ref{seriesCA}) and (\ref{seriesYC}), we can see that the engine operating at maximum efficient power function is more efficient than the engine  at maximum power.	We plot 	Eqs. (\ref{seriesCA}) and (\ref{seriesYC}) in Fig. 5.
 	\begin{figure}
		\begin{center}
			\includegraphics[width=8.6cm]{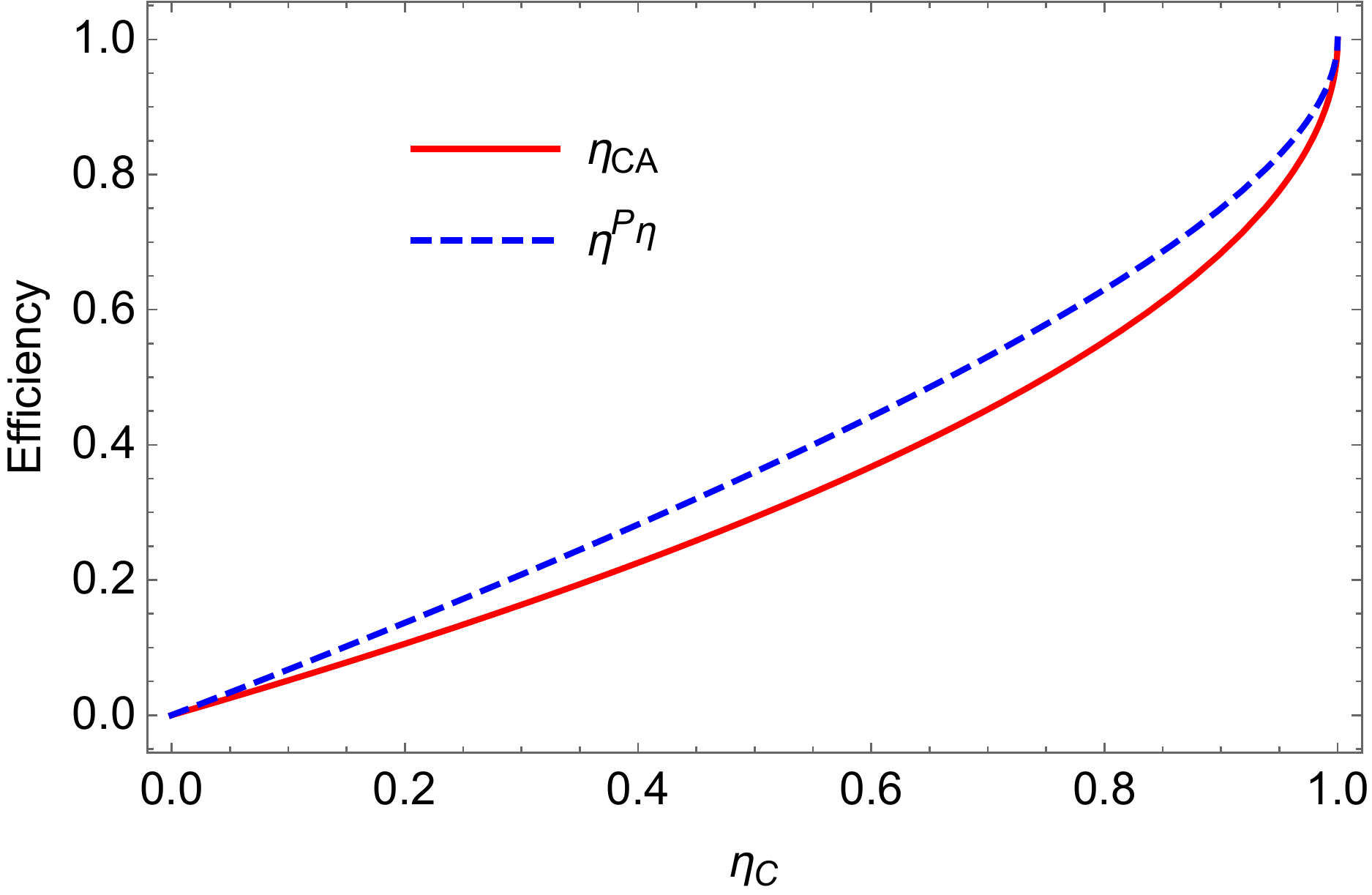}
		\end{center}
		\caption{Plots of  efficiency at maximum efficient power, Eq. (\ref{seriesYC}) (dashed blue curve), and efficiency at maximum power, Eq. (\ref{seriesCA}) (solid red curve), as a function of Carnot efficiency $\eta_C$ for the endoreversible quantum heat engine operating in the high temperature regime.}
	\end{figure}
\subsection{Fractional loss of power}
In this section, we compare the fractional loss of power due to entropy production in our quantum endoreversible heat engine operating at maximum efficient power function to that of operating at maximum power.    
Power loss due to entropy production is given by: $P_{\rm lost}=T_2\dot{S}_{\rm tot}=\dot{Q}_c-(1-\eta_C)\dot{Q}_h$.
Further using the definitions of power output $P=\dot{Q}_h-\dot{Q}_c$ and efficiency $\eta=P/\dot{Q}_h$, the ratio
of power loss to power output can be derived as \cite{MePRR}:
\begin{equation}
R \equiv \frac{P_{\rm lost}}{P} = \frac{\eta_C}{\eta}-1. \label{ratioplost}
\end{equation}
First, we will discuss the fractional power loss in the engine operating under the conditions of maximum efficient power function. Using Eq. (\ref{seriesYC}) in Eq. (\ref{ratioplost}), we have
\begin{equation}
R^{P_\eta} =  \frac{1}{4}\big[\eta_C + \sqrt{(1-\eta_C)(9-\eta_C)}-1\big], \label{plostEP}
\end{equation}
Similarly, we can obtain the corresponding expression for the engine at maximum power by using Eq. (\ref{seriesCA}) 	 in Eq. (\ref{ratioplost}):
\begin{equation}
R^{P} = \sqrt{1-\eta_C}. \label{plostP}
\end{equation}		
We plot	Eqs. (\ref{plostEP})	and (\ref{plostP})	in Fig. 6 as a function of Carnot efficiency $\eta_C$.
	\begin{figure} [h]
		\begin{center}
			\includegraphics[width=8.6cm]{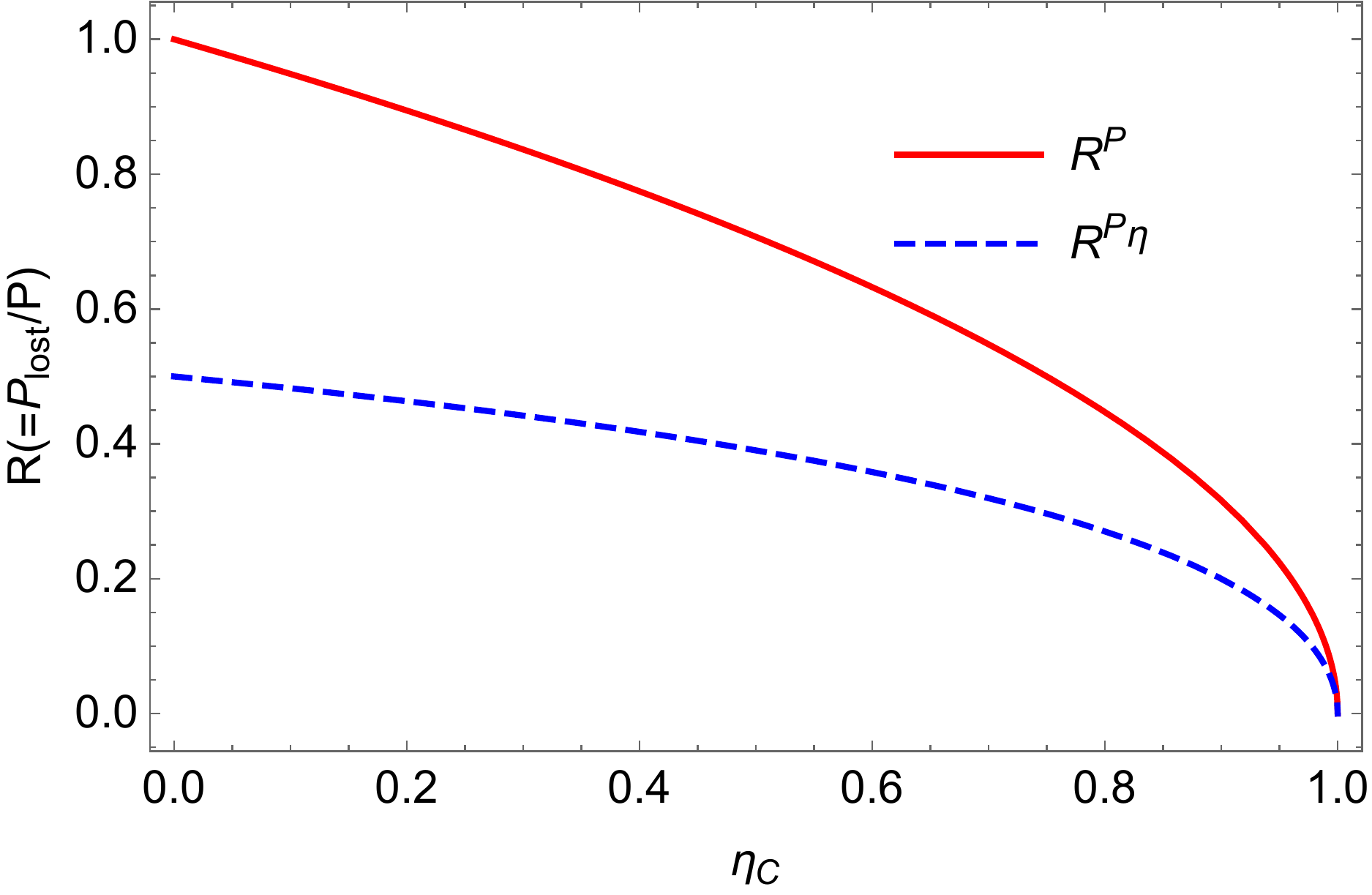}
		\end{center}
		\caption{Comparison of fractional loss of power for two different optimization functions: efficient power function and power output. The lower-lying dashed blue curve (Eq. (\ref{plostEP})) represents the case when efficient power function is optimized whereas the upper lying curve  (Eq. (\ref{plostP})) represents the  case when  power output is optimized. } 
\end{figure}
It is clear from Fig. 6 that the 	fractional loss of power is much lower 	when our engine operates in maximum efficient power regime (dashed blue curve in Fig. 6) as compared to the case when our engine operates in maximum power regime (solid red curve in Fig. 6), which in turn implies that engines at maximum efficient power wastes less amount of power thus keeping a check on the environmental pollution and fuel wastage. Thus, we can conclude that the efficient power is a good objective function from the point of view of environment and fuel conservation.

\subsection{Ratio of power at maximum efficient power to maximum power}
Besides the  fractional loss of power, another useful quantity to calculate  the ratio ($R'$) of power at maximum efficient power to maximum power.  The expression for the maximal power can be obtained by substituting Eq. (\ref{seriesCA}) in Eq. (\ref{auxxpower}). We have
\begin{equation}
P^{\rm max}_{\rm HT} = \frac{k_{1} k_{2} }{k_{1}+k_{2}} T_h \eta \left(1 - \frac{1-\eta_C}{1-\eta}\right) = (1-\sqrt{1-\eta_C})^2. \label{RR1}
\end{equation}
Similarly, we can obtain the expression for power  at maximum efficient power function by substituting Eq. (\ref{seriesYC}) in Eq. (\ref{auxxpower}):
\begin{equation}
P^{\rm EP}_{\rm HT} = \frac{1}{4} \left(-3 \sqrt{\left(1-\eta _C\right) \left(9-\eta _C\right)}-5 \eta _C+9\right). \label{RR2}
\end{equation}
Dividing Eq. (\ref{RR2}) with Eq. (\ref{RR1}), we obtain the following expression for the ratio   of power at maximum efficient power to maximum power
\begin{equation}
R' = \frac{ \left(-3 \sqrt{\left(1-\eta _C\right) \left(9-\eta _C\right)}-5 \eta _C+9\right)}{4 (1-\sqrt{1-\eta_C})^2}. \label{ratioo}
\end{equation}
For $\eta_C\rightarrow 0$, the ratio $R'=P^{\rm EP}_{\rm HT} /P^{\rm max}_{\rm HT}=8/9$, which indicates that at least 88.89$\%$ of the maximum power is produced when our engine operates in the maximum efficient power regime, which is a considerable amount since  the power loss in maximum efficient power regime is regime is  exactly 1/2 (for $\eta_C\rightarrow 0$) of the case when the engine operates at maximum power.  
\begin{figure}   
 \begin{center}
\includegraphics[width=8.6cm]{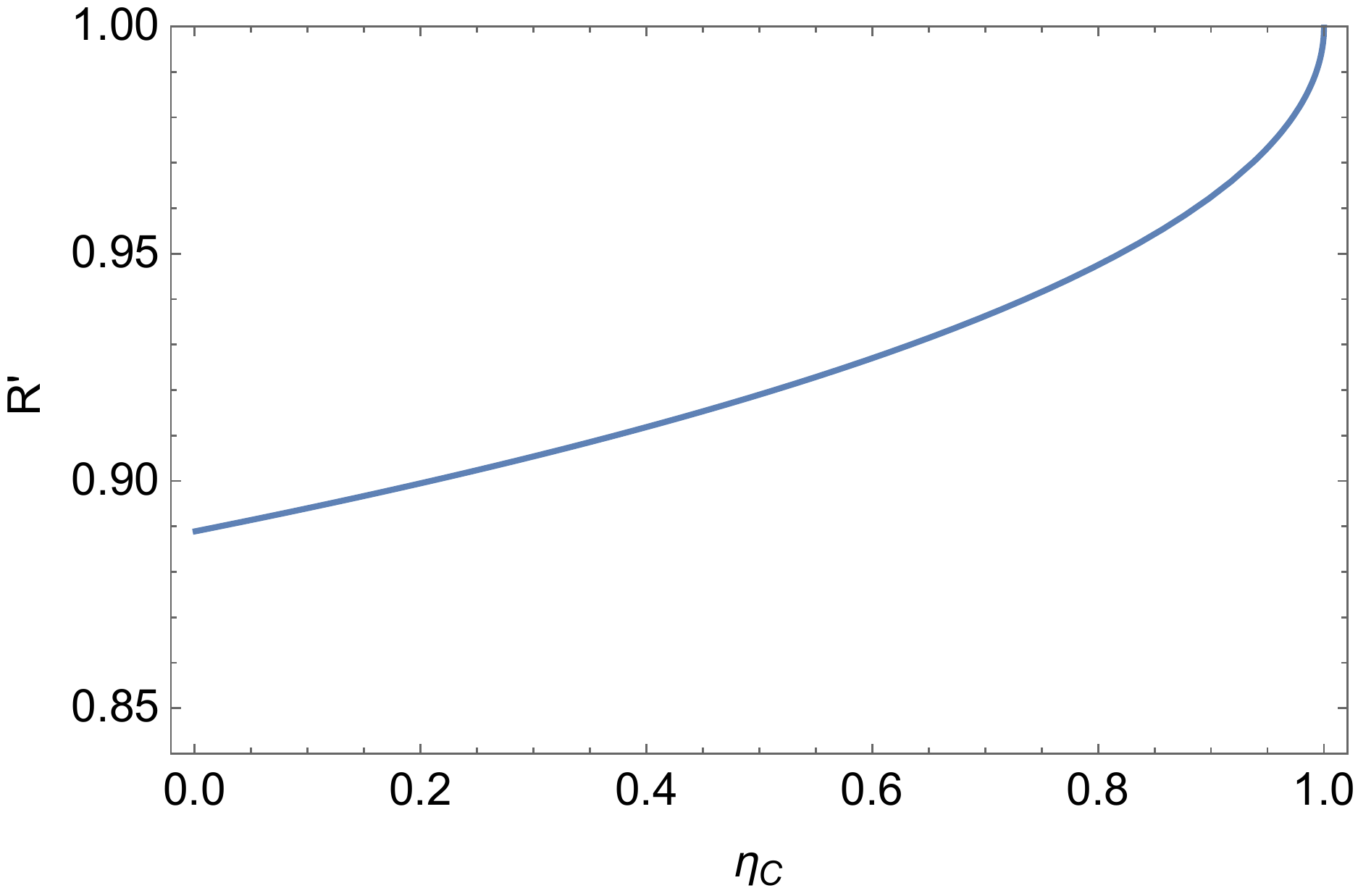}
 \end{center}
\caption{ Ratio $R'$ (Eq. (\ref{ratioo})) of the power output at maximum efficient power (Eq. (\ref{RR2}))    to the maximum power (Eq. (\ref{RR1})) as a function of Carnot efficiency $\eta_C$.}
\label{ratioPEP}
\end{figure}

\section{stochastic simulations}
We studied the thermal fluctuations for quantum endoreversible engine \cite{Fernandez2022}.  For stochastic simulations we choose three reference efficiencies $\eta_{0} = 0.45$ (where both P and ${P_{\eta}} $ are positive and have small values), $\eta_0=\eta^{P}$ (efficiency at   maximum power) and $\eta_0=\eta^{P_{\eta}}$ (efficiency at maximum efficient power).  $\eta$ is varied randomly around  these reference frequencies and we calculate the change in power ($\Delta P$) of the engine for each case. The graphs shows that the power $P$ oscillates randomly around mean $P$ for each case (see Fig. 8). The oscillations are largest for the case with $\eta_0=0.45$. However, we are more interested in comparing the the cases  when chosen reference efficiencies are $\eta_0=\eta^{P}$  and $\eta_0=\eta^{P_{\eta}}$. We find that the engine is subjected to relatively larger power fluctuations (orange colored oscillations in Fig. 8)  when operating in the maximum efficient power regime as compared to the one at maximum power. Thus we conclude that the engine operating in the maximum power regime is the more stable one.

In our  simulations, we restrict the change in $\eta$  to the range [0.99$\eta_{0}$,1.01$\eta_{0}$] for all three cases. Table 1 provides numerical values  of the dispersion in power for each case.

				\begin{figure}
					\begin{center}
						\includegraphics[width=9.4cm]{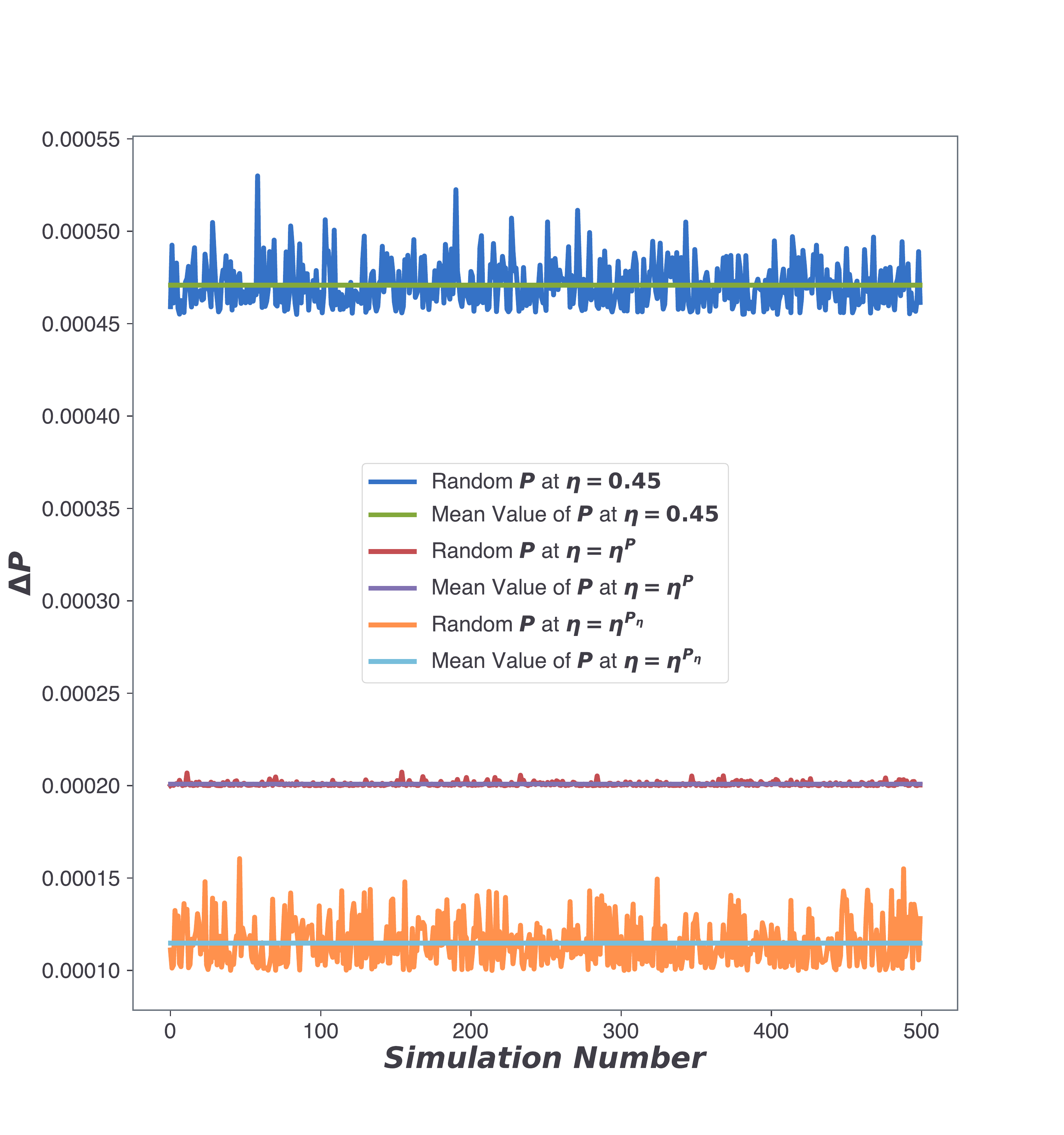}
						\end{center}
					\caption{ Fluctuations in power, $\Delta P$, of the engine for the choice of three different reference efficiencies: $\eta_{0}$ = 0.45, $\eta_0=\eta^{P}$ and  $\eta_0=\eta^{P_{\eta}}$.  The other parameters are fixed at constant values, $E_1=10$, $E_2=6$, $\Gamma_{h}=\Gamma_{1}=0.01$, $\Gamma_{c}=\Gamma_{2}=0.001$, 	 $T_{h}=10$, $T_{c}=1$. .  } 
				\end{figure}
					\begin{table}[h!]
			\begin{center}
				\caption{Mean values and dispersions for the $\bigtriangleup P$ and $\bigtriangleup P_{\eta}$ at
					three different reference efficiencies. The data have been obtained by 500 simulations. }
				\label{tab:table1}
				\begin{tabular}{|c|c|c|c|c|}
					\hline
				Ref.Effi.	&$< \bigtriangleup P >$  & $\sigma_{P}$  &$<\bigtriangleup P^{\eta}>$  &$\sigma_{P_{\eta}}$  \\
					\hline
					$ 0.45$	&$1.51 \times 10^{-5} $  & $1.18 \times 10^{-5} $ & $1.56 \times 10^{-5} $ & $1.18 \times 10^{-5} $ \\
					\hline
					$ \eta^{P}$	&$9.99 \times 10^{-7} $ &$1.23 \times 10^{-6} $  & $9.49 \times 10^{-6} $ &  $7.35 \times 10^{-6} $\\
					\hline
					$ \eta^{P_{\eta}} $	& $1.42 \times 10^{-5} $ & $1.20 \times 10^{-5} $ & $1.07 \times 10^{-6} $ & $1.39 \times 10^{-6} $ \\
					\hline
				\end{tabular}
			\end{center}
		\end{table}

		\section{Conclusion}
		In this work, we have investigated the optimal performance of  an endoreversible quantum dot heat engine in which irreversible heat transfer between the baths and the working fluid is mediated via qubits. In order to operate the engine in a regime paying equal attention to power production as well as efficiency, we  chose  efficient power function as our objective function to optimize. First, we numerically studied the optimization of efficient power function by fixing energy gap $E_1$ of qubit 1 or by fixing energy gap $E_2$ of qubit 2. Here we learned that unlike classical endoreversible engine, power and efficiency of the quantum endoreversible heat engine are regulated by the microproperties of the qubits. Then we studied the optimal performance of the engine under consideration in the high temperature regime and obtained analytic expression for the efficient at maximum efficient power function. Additionally, we compared the optimal performance of the quantum endoreversible engine engine operating at maximum efficient power  to that of operating at maximum power. We showed that the fraction loss of power due to entropy production is considerably for the engine at maximum efficient power while at the same time it produces at least 88.89\% of the maximum power output.	Finally, we also investigated the stability of our engine against thermal fluctuations. We found that the engine operating under the conditions of maximum power is subjected to less fluctuations than the engine operating in the maximum efficient power function, thus rendering it more stable.

		\section{Acknowledgements}
		The author gratefully acknowledges  the name of Dr. K P Sharma, for his great contribution in resource management.
					\bibliography{endo_quantum.bib}
	\bibliographystyle{apsrev4-2}
		\end{document}